# Deep Neural Networks integrating genomics and histopathological images for predicting stages and survival time-to-event in colon cancer

Olalekan Ogundipe[1], Zeyneb Kurt[1*], Wai Lok Woo[1]

[1]Department of Computer and Information Sciences, University of Northumbria, Newcastle Upon Tyne, UK

*Corresponding author: zeyneb.kurt@northumbria.ac.uk

**Abstract**

Motivation: There exist unexplained diverse variation within the predefined colon cancer stages using only features either from genomics or histopathological whole slide images as prognostic factors. Unraveling this variation will bring about improved in staging and treatment outcome, hence motivated by the advancement of Deep Neural Network (DNN) libraries and different structures and factors within some genomic dataset, we aggregate atypia patterns in histopathological images with diverse carcinogenic expression from mRNA, miRNA and DNA Methylation as an integrative input source into an ensemble deep neural network for colon cancer stages classification, and samples stratification into low or high risk survival groups.

Results: The results of our Ensemble Deep Convolutional Neural Network (EDCNN) model show an improved performance in stages classification on the integrated dataset. The fused input features return Area under curve – Receiver Operating Characteristic curve (AUC-ROC) of 95.21% compared with AUC-ROC of 71.09% and 67.98% obtained when only genomics and images features are used for the stage's classification, respectively. Also, the extracted features were used to split the patients into low or high-risk survival groups. Among the 2,548 fused features, 1,695 features showed a statistically significant survival probability differences between the two risk groups defined by the extracted features.

**Availability and Implementation:** https://github.com/Ogundipe-L/EDCNN

**1 Introduction**

There are over 67 cancer's primary site in the human body, among which are brain, breast, colon and several other cancer types that have been identified and detected in various parts of the body, with the lung having the highest cases and the colon rank fourth. According to World Health Organization 2020 report, cancer is the leading cause of death accounting for over ten million deaths. Available research tools and dataset provided by cancer research community for analyzing causes and treatment of cancer types are genomics and histopathological whole slide images datasets among others, each consisting high-dimensional features embedding the complex pathological pattern of each cancer stages. The complex features comprising the tumor size, the extent at which the cancer has spread to nearby lymph node, and whether the cancer has metastasized, which is the extent the cancer has spread to other parts of the body from the primary tumor still require the development of an efficient and effective model for accurate and improve classification and effective treatment management.

Colon cancer is a malignant tumor of the large intestine or of the rectum which affects both males and females irrespective of age group (M. Y. Lu et al., 2021; Pacal et al., 2020; Shapcott et al., 2019; Song et al., 2018). This cancer could occur by genetic mutations or through hereditary in an individual. Hundreds of million cases have been diagnosed every year and tens of thousands death reported globally every year (Gupta et al., 2021). In terms of mortality rate, colon cancer in ranked second (Munir et al., 2019; Wulczyn et al., 2021). Cancer diagnosis can be classified into four stages ranging from I to IV where stage I implies that the cancer is in early stages and IV indicates that the cancer is in advanced stages (https://www.cancercenter.com/cancer-types/colorectal-cancer/stages).

One of the most challenging area in cancer research is staging, which is an indicator of the patient most likely outcome, life expectance and chances of cure (Abdelhafiz et al., 2019; Benhammou et al., 2020; Coebergh Van Den Braak et al., 2020; D'Amour et al., 2020). Researchers studying colon cancer are working on various ways to unravel the means of prevention, treatment and timely detection of the disease to reduce the mortality and incidence rate and improving the quality of life for people infected with colorectal cancer (Chaudhary et al., 2019; Xie et al., 2017). Deep learning computational framework has been deployed in the study and analysis of biospecimen data ranging from features encoding and extraction, integration and transformation of heterogeneous biological data (Kather, Pearson, et al., 2019; Komura & Ishikawa, 2018; Xia et al., 2019). Most of previous studies only concentrated on unimodal data feature study, and some that are based on multimodal data are either purely clinical trials (Gulum et al., 2021; Mobadersany et al., 2018; Schmauch et al., n.d.), or uses linear models and eigengenes to extract features as the baseline framework (Cheng et al., 2017) , or are earlier deep learning framework (Kather, Krisam, et al., 2019). However, our study relies on bimodal fusion of features from genomics (DNA methylation, mRNA, and miRNA) and slide images cumulating into quadruple datasets for tumor stages examination using ensemble based deep learning framework. The analysis and result of our proposed model shows an improved and effective cancer stages classification and survival risk group stratification based on fused features from genomics and histopathological whole slide images.

## 2 Materials and Methodology

2.1 Samples

The data used in this study was downloaded from The Cancer Genome Atlas (TCGA) using the module A package of TCGA-Assembler function and cut across six colon cancer disease types (adenomas and adenocarcinomas, cystic, mucinous and serous neoplasms, complex epithelial neoplasms, mature b-cell lymphomas, epithelial neoplasms and lipomatous neoplasms). It initially comprises of 448 samples of clinical and demographics data, 328 samples of mRNA expression, 261 samples of miRNA expression, and 353 samples of DNA Methylation. After a series of sorting and filtering operations on data downloaded from TCGA, we are left with only 177 patients having data across the four datasets (histopathological images, mRNA-seq, miRNA and DNA methylation). A supervised method of deep neural network learning was designed on the integrated dataset with the ajcc-pathologic-tumor-stage column of the clinical data used as the label in training the proposed model. Table 1. Shows the distribution of the data used across the four cancer's stage in this study.

Table 1. Patient data distribution across images and genomics

| | Whole slide images | | Genomics sample size | | |
|---|---|---|---|---|---|
| stages | sample size | Number of Tiles | mRNA | miRNA | DNA Methylation |
| I | 30 | 27,358 | 30 | 30 | 30 |
| II | 67 | 47,397 | 67 | 67 | 67 |
| III | 54 | 21,492 | 54 | 54 | 54 |
| IV | 26 | 15,914 | 26 | 26 | 26 |
| | 177 | 112,161 | 177 | 177 | 177 |

2.2 Proposed Ensemble Deep Convolution Neural Network (EDCNN) framework

For effective stages classification and clinical patient risk stratification, we used an ensemble of deep neural network for extraction and fusion of biological data of different molecular representation and correlation structure. The autoencoder function for features extraction from structure genomics and resNet50 for features extraction from histopathological images. Figure. 1 is our proposed ensemble DCNN which encapsulates the following steps: (i) Figure 1(a) we merge features from mRNA, miRNA and DNA Methylation and abstract high-level feature space where the shared representations across different biology molecular datasets are fused through sparse autoencoder (AE). (ii) Figure. 1(b) salient features extraction from histopathological whole slide image tiles. (iii) Figure 1(c) Integration/fusion of features in steps (i) and (ii) followed by colon cancer stages classification and risk stratification by CNN with fully connected layer. The detail implementation of our proposed framework will be covered in the following subsection.

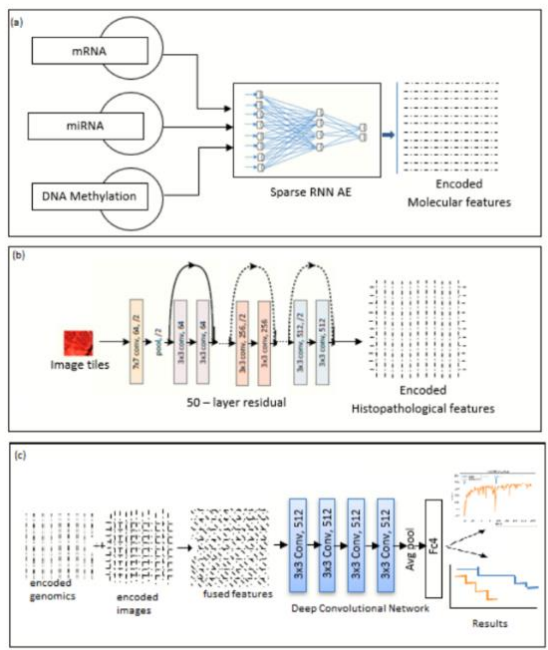

*Figure 1 The proposed EDCNN model with three major steps (a) extraction of salient features from concatenated biological genes of mRNA, miRNA and DNA- Methylation (b)feature extraction from histopathological images (c) CNN stages prediction with fused features*

2.3 Ensemble Deep Convolution Neural Network (EDCNN) implementation

To implement the first two steps of the proposed EDCNN which are features extraction processes, we choose two network models which are sparse autoencoder (AE) (Lopez Pinaya et al., 2019) and ResNet50 (He, n.d.). The sparse autoencoder is adopted for joint encoding merged features from mRNA, miRNA and DNA Methylation using combinations of non-linear functions. The resulting encoded genomics features is embedded with robust phenotypic nomenclature more definitive for the cancer staging and survival risk stratification. The second model, ResNet50 - a 50 layers residual deep neural network was also used for features extraction of histopathological images dataset. The implementation procedure for feature extractions through AE and ResNet50 follows:

We input 18,364 merged features from miRNA, mRNA and DNA Methylation into autoencoder (AE) neural network design specifically to encode its input. The AE leverages a scalable hyperparameter optimization framework that searches the AE space for best hyperparameter values for a sequence of bottleneck encoding features that will ensure optimal performance of the AE network for the classification and stratification purposes. As described in (Lopez Pinaya et al., 2019) the EA learns the mapping $Y: \Re^n \to \Re^p$ (encoder) and $Z: \Re^n \to \Re^p$ (decoder) that satisfy

$$\arg\min_{X,Y} E[\Delta(a, ZoX(a))]. \tag{1}$$

Where *a is the inputs to the first layer of the AE, $ZoX(a)$* are mappings to be fit by AE neural network, *E* is the expectation over the distribution of *a*, and $\Delta$ is the encoding loss function, which measures the distance between the output of the decoder and the input. The encoding loss function is usually set to be the $L_2$-norm. We reduce the construction error on the bottleneck layer by adding the sparse regularization. The sparse regularization controls the bias-variance tradeoff that is encounter by the network. Hence, the AE optimization objective function becomes:

$$\arg\min_{X,Y} E[\Delta(a, ZoX(a))] + \lambda \sum_i |y_i| \tag{2}$$

Where $y_i$ is the activation at the *ith* hidden layer and *i* iterates over all the hidden activation. We extend AE network and measure the bottleneck layer that returns highest accuracy on decoding the reconstructed features. The solution to our encoding problem now becomes

$$Y: \mathfrak{R}^{18364} \to \mathfrak{R}^{500} \tag{3}$$

Which implies that $p = 500$ is the number of features extracted from the 18,364 merged input features.

For feature extraction from histopathological images, our EDCNN rely on ResNet50. ResNet50 is the 50 layers of ResNet framework proposed by (He, n.d.). ResNet is a residual learning mechanism embedded within the deep Convolutional Neural Network (CNN) to make it to be more effective. The residual training mechanism is designed to handle dual problems of vanishing gradients and degradation of training accuracy. Traditionally, a CNN contains three types of layers: convolution, pooling and fully connected layers.

As described in (S. Lu et al., 2020) Convolution operation generates feature maps from the input by a set of filters with trainable weights. Given an image *I* in size of (*M*, *N*) and a filter *F* in size of (*p*, *q*), the convolution formula can be expressed as

$$conv = (I * F)(x, y) = \sum_M \sum_N I(x-p, y-q) F(p, q) \tag{4}$$

Pooling is a simple but effective operation, and pooling layers are placed after convolution layers. Pooling generates feature maps with a local perceptive field. Pooling layers extract main features from the input and reduce the dimension, which helps accelerate the training and improve the generalization ability.

The idea is that network non-linear layers asymptotically approximate the residual function. For instance, $f(x)$ denotes the learned mapping of several layers, then it is equivalent to train these layers to approximate the residual function: $r(x) = f(x) - x$. So, in residual learning, the target function becomes:

$$f(x) = r(x) + x \tag{5}$$

These implementation as denoted by equation (5) ensures no generation of extra parameter and the computation complexity of the classifier is kept constant.

2.4 Framework for mapping features extracted from histopathological images and genomics

Methods for combining vectors representation from histopathological and genomics include element-wise product or sum, Multimodal Compact Bilinear pooling (MCB) (Fukui et al., 2016) and concatenation. Bilinear pooling computes the outer product between two vectors, which allows, in contrast to element-wise product, a multiplicative interaction between all elements of both vectors. On the other hand, the concatenation method linked the two representations in series. Individually, each type of data adopted carries different mutated and complex topological outlay of cancerous tumor, each with limited amount of biology carcinogenic factors and combining them would provide a more complementary and comprehensive prediction features. We aggregate the encoded features in each data by concatenating extracted features from the genomics and histopathological images to create new multi-dimension features for sample represented as:

$$X(conc[k^n \oplus h^m]) \rightarrow Y(l^{n \oplus m}) \tag{6}$$

2.5 Training, Evaluation and Testing

The unimodal and multi-modal features are trained under the same experimental condition. The classifier was trained trice, the first and the second training was done with encoded genomics and histopathological image data as input, while the third training used input from the integrated features of the encoded genomics and histopathological image. The training and validation sample data comprises 80% of the entire dataset, while the remaining 20% is used for testing. To ensure correction for the imbalance distribution of cancer stages in the sample data, we use stratified 5-fold cross-validation to ensure relative class frequencies is approximately preserved in the training, validation, and testing fold.

2.6 Classification model

We hypothesized that synthesizing and aggregating extracted features from histopathological images and genomics as input data into CNN for colon cancer stages prediction and risk stratification will lead into an improvement and more accuracy in prediction and survival risk stratification. We test the validity of our hypothesis after reducing the dimension and consequently learn a higher-dimensional and compressed representation from the genomics and histopathological images as shown in Figure. 1(a) and Figure. 1(b). We used T-distributed Stochastic Neighbor Embedding (t-SNE) to project and visualized each set of extracted and integrated extracted features. t-SNE works by converting similarities between data points to joint probabilities and minimize the Kullback-Leibler divergence between the joint probabilities of the low-dimensional embedding and the high-dimensional data. Thereafter, the concatenated features are then used as an input into a convolution neural network for colon cancer stages prediction and for risk stratification. The CNN classifier comprises of three different layers: the input layer, a middle dense layer, and the output layer. The classifier is parameterized with categorical crossentropy loss function, Adam optimizer (a replacement optimization algorithm for stochastic gradient descent for training deep learning models) and metric accuracy. The training and validation fitting is set to run for 200 epochs. The experiment was designed and implemented on a server with ubuntu operating system fitted with 3 sets of NVIDIA GeForce GTX 1080 Ti GPU devices with a memory size of 10410 MB per GPU device. The entire code used during the implementation are python based on TensorFlow, Pandas, and scikit-learn libraries. Our proposed ensemble model is used to train each group of features. Two unimodal classification, one based on image features and the other based on genomics features with the third classification based on fused features from images and genomics, and we measure the accuracy of DNN cancer stage prediction with Area Under the Curve Receiver Operating Characteristic (AUC ROC) metrics.

## 3. Results

3.1 Extracted features from genomics data types

After the preprocessing steps on mRNA, miRNA and DNA Methylation, we combine and extract features from these three genomics datasets with interrelated biological and transcriptional function and used the joint genomics extracted features obtained from bottleneck of sparse autoencoder for training the DNN model. The accuracy and performance metrics of our model obtained. Projection and visualization of the extracted genomics features from high dimensional space embedded into low dimension latent space was performed through the t-SNE visual representation function Figure. 2(A) showing insight into extracted features from the three-genomics data type.

Figure 3(A) shows the model accuracy based on training with the joint genomics features. The result shows that at the early stage of the training the model accuracy was very high at approximately 95% within the first 30 epochs, and as the training times increases with increase in the number of epochs, the model accuracy converges to 82% and stabilize to approximately 72%. The testing accuracy and prognostic performance of the model on never seen (genomics features) data as reflected in the confusion matrix - Figure. 4(A) and the AUC-ROC Figure. 4(D) is as summarized in the second column of Table 3. It shows that our model predicts the colon cancer stage of class I with 74% accuracy, class II with 71% accuracy, class III with 75% accuracy and class IV with 65% accuracy when the input into the model are features extracted from genomics dataset.

3.2 Features from histopathological images

Insight into abstract level extraneous features from 224-by-224 histopathological image tiles is shown in Figure 2(B). showing stages stratified distribution across four cancer stages. Aggregating tiles level features into high dimensional extraneous features is achieved through the global pooling layer of ResNet50 pretrained model. 2,048 features are extracted from each tile through transfer learning and feature extraction mechanism. Figure 2(B) is the t-SNE visual representation of the abstract level of the extracted histopathological images showing the four classes of colon cancer stages.

Figure 3(B) shows the result of model accuracy based on training with the histopathological image features. It shows that at the early stage of the training, the model accuracy was low at approximately 62% within the first few epochs, and as the training epochs progress, the model accuracy also increases and stabilize to approximately 78%. The testing accuracy and prognostic performance of the model on never seen (mage tiles) data as reflected in the confusion matrix - Figure. 4(B) and the AUC-ROC Figure. 4(E) is as summarized in the third column of Table 3. Our model predicts the colon cancer stage of class I with 71% accuracy, class II with 67% accuracy, class III with 66% accuracy and class IV with 68% accuracy when the input into the model are features extracted from histopathological image dataset.

3.3 Integrated image and genomics features

Figure 2(C) shows embedded latent space of the fused extracted features clearly mapping each sample to separate cluster cancer stages. The t-SNE is more sensitive to local structure in the fused features than in the individual unimodal features. The t-SNE function converts affinities of data points in the unimodal and multimodal to probabilities. The affinities in the original space are represented by Gaussian joint probabilities and the affinities in the embedded space are represented by Student's t-distributions revealing data that lie in different cancer stage clusters.

Figure 3(C) shows the result of model accuracy based on training with the fused features. The result shows that at the early stage of the training, the model accuracy increases gradually from at approximately 93% within the first few epochs, and as the training epochs increase, the model accuracy stabilize to approximately 98%. The testing accuracy and prognostic performance of the model on never seen (integrated) data as reflected in the confusion matrix - Figure. 4(C) and the AUC-ROC Figure. 4(F) is as summarized in the fourth column of Table 3. The result shows that our model predict the colon cancer stage of class I with 91% accuracy, class II with 95% accuracy, class III with 96% accuracy and class IV with 99% accuracy when the input into the model is fused extracted features from genomics and histopathological image dataset.

Figure 2(C) is a visual representation of the outcome confirming our initial hypothesis that integrating features extracted from histopathological images and genomics will lead to an improve and accuracy in stage classification and survival risk analysis. This outcome follows the pattern and the semblance of the American Joint Committee on Cancer (AJCC) TNM staging system

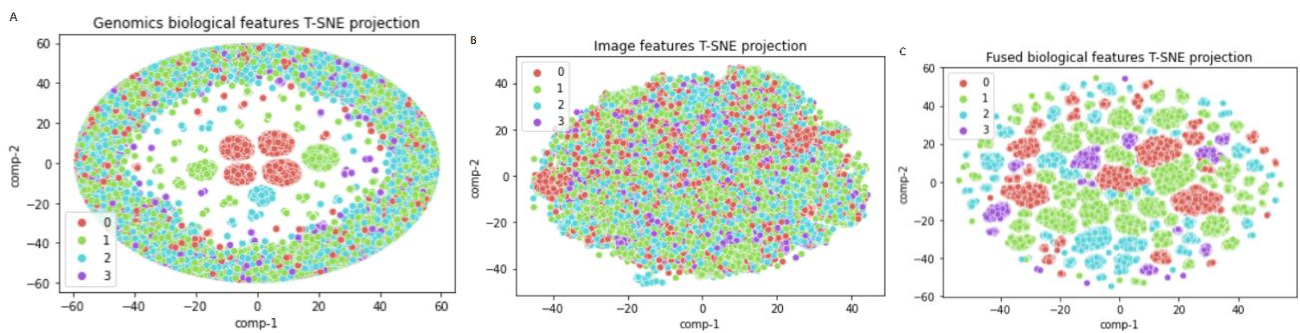

*Figure 2 t-SNE Visualization of sample extracted features from histopathological and genomics mapped into four cancer stages. (A) embedded visualization from extracted genomics features mapped into cancer stages (B) embedded visualization from samples features*

*extracted from histopathological images mapped into cancer stages (C) embedded space of samples fused genomics and histopathological extracted features mapped into cancer stages.*

3.4 Integrated Latent Space and survival analysis

We analyze the proportion of the fused 2,548 extracted features that will group the patients into low and high-risk survival groups necessary for assessing the survival outcome. The survival analysis was built around three vectors representation. The first vector been the fused features and the other two vectors - vital status and last contact days are from clinical record of patient with colon cancer disease. To achieve our set objectives, the median value of each extracted features was estimated, and sample value under each feature classified as either belonging to low or high-risk survival group compared with its median value. The Kaplan-Meier estimator was used to estimate and visualize the difference between the low and high-risk survival sample groups under each feature individually and compare-survival algorithm of the scikit-survival library is used to compute the p-values related to statistical significance between the low and high-risk groups. The Kaplan-Meier method is an estimator for analyzing time-to-event data whose survival-time covariates are right censored, while the compare-survival function is the K-sample log-rank hypothesis test of identical survival functions. It compares the pooled hazard rate with each group-specific hazard rate. The alternative hypothesis is that the hazard ratio of at least one group differs from the others at some time. To control for the probability of committing a type I error among the n=2548 statistical test, we adopt Bonferroni Correction (BC) for adjusting the alpha level value which was originally set to 0.05. The BC is given as:

$$\alpha_{new} \leq \frac{\alpha_{original}}{n} \quad (5)$$

Where $\alpha_{newl}$ is the adjusted alpha level, $\alpha_{original}$ is the original alpha level and n is total number of statistical tests perform. This translates into a situation that we reject the null hypothesis of each test

within the multiple statistical tests only when the raw p-value is less than

$$\alpha_{new} \leq \alpha_{original}/n = 0.05/2548 = 1.962e^{-5}.$$

3.5 Survival risk stratification with fused features

The Kaplan-Meier estimator of 10 most significant extracted features showing the survival functions of patient in low and high-risk survival, Figure 5 gives the visual representation of the relationship between time and the probability of patient in low or high-risk class surviving beyond a given time point. As shown in Figure. 5, the wide gap between the two functions is an indicator that we can confidently argue that most of the extracted features conveniently group the samples into low or high-survival class. The result of *p*-values from log-rank Statistical test supporting that majority of the extracted features was able to stratify the patients into low and high-risk survival groups is as shown in Table 4. The features with the top 10 most significant real p-value confirmed the statistically significant difference between the low and high-risk survival groups. The significance of the estimated real p-values supports the evidence that the fused features perform better in predicting colon cancer stages and stratifying samples into low and high-risk survival groups required and necessary for clinical prognosis.

Table 2. Training, validation, and testing accuracy based on each category of features.

| accuracy | Classification based on images | Classification based on genomics | Classification based on fused features |
|---|---|---|---|
| Training | 78.4% | 72.7% | 99.8% |
| Validation | 74.7% | 76.0% | 98.1% |
| Testing | 53.3% | 67.1% | 91.5% |

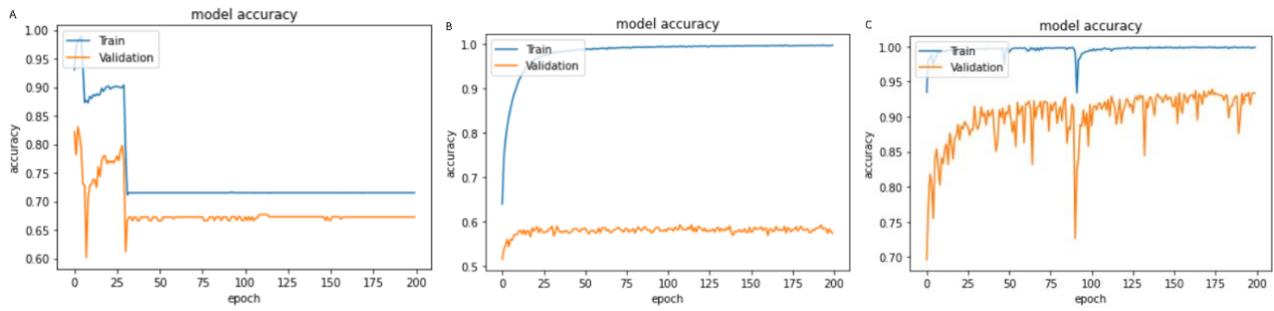

*Figure 3 The model accuracy based on: (A) genomics data, (B) image, and (C) fused features*

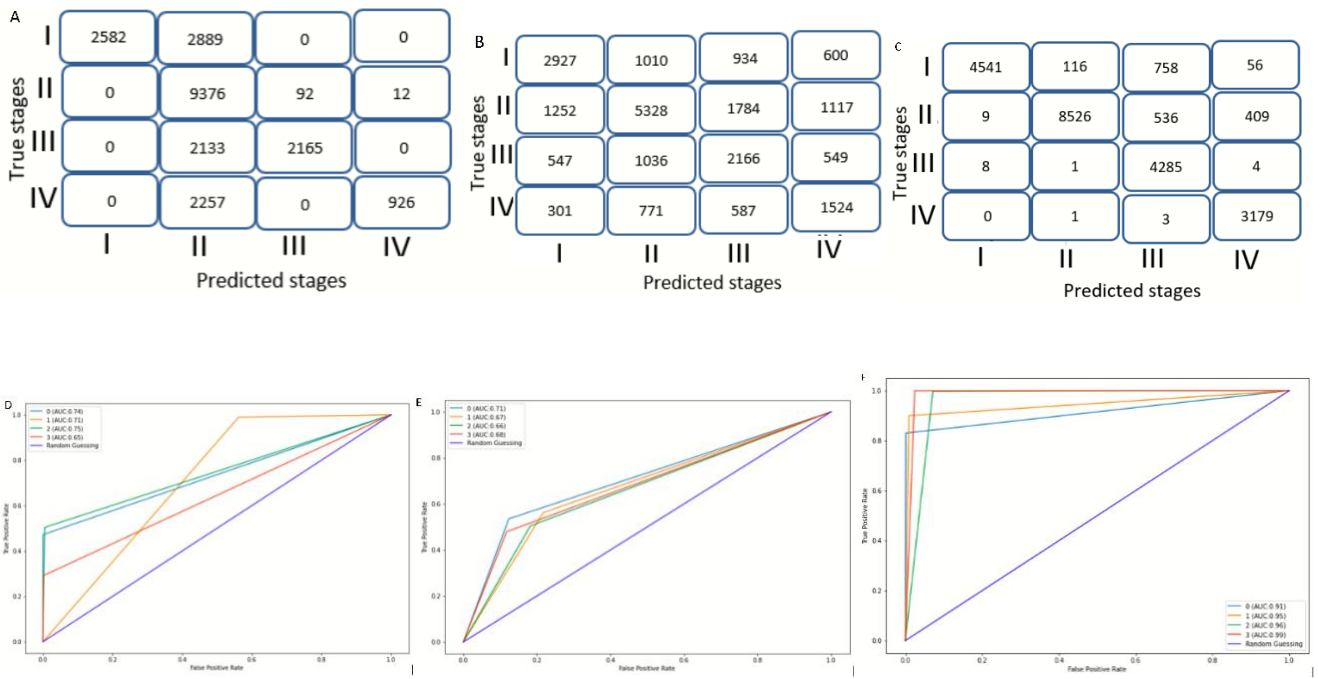

*Figure 4 Colon cancer stages classification results: Confusion matrix based on (A) genomics, (B) on image, (C) based on fused features. AUC-ROC curves for: (D) genomics (E) images, (F) fused features*

Table 3. AUC-ROC estimate for each cancer stages under each category of features

| Cancer stages | AUC based on features from genomics (% accuracy) | AUC based on features from images (% accuracy) | AUC based on features from wsi + genomics (% accuracy) |
|---|---|---|---|
| I | 74.3 | 71.1 | 91.4 |
| II | 71.0 | 67.4 | 95.0 |
| III | 75.1 | 66.5 | 96.2 |
| IV | 65.7 | 68.3 | 99.3 |

Table 4. The first 10 features from the fused features with significant p-values characterizing the low-high risk samples stratification in Kaplan Meier curve

| extracted features | the number of the samples in low-risk group | the number of the samples in high-risk group | p-value |
|---|---|---|---|
| v2329 | 47766 | 51042 | 1.05E-113 |
| v541 | 49403 | 49405 | 1.67E-113 |
| v2391 | 49432 | 49376 | 1.02E-105 |
| v870 | 49404 | 49404 | 7.60E-102 |
| v2297 | 50689 | 48119 | 2.01E-86 |
| v94 | 49404 | 49404 | 4.53E-85 |
| v1 | 49404 | 49404 | 5.90E-84 |
| v174 | 49404 | 49404 | 9.07E-80 |
| v333 | 49404 | 49404 | 1.51E-77 |
| v637 | 49404 | 49404 | 1.44E-76 |

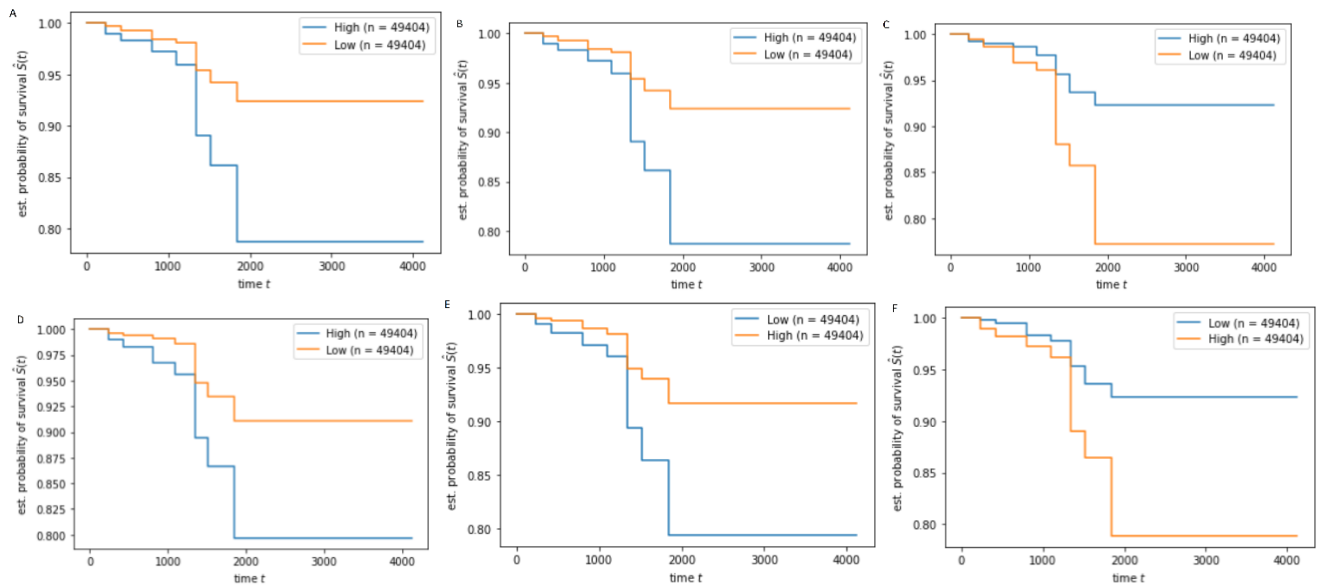

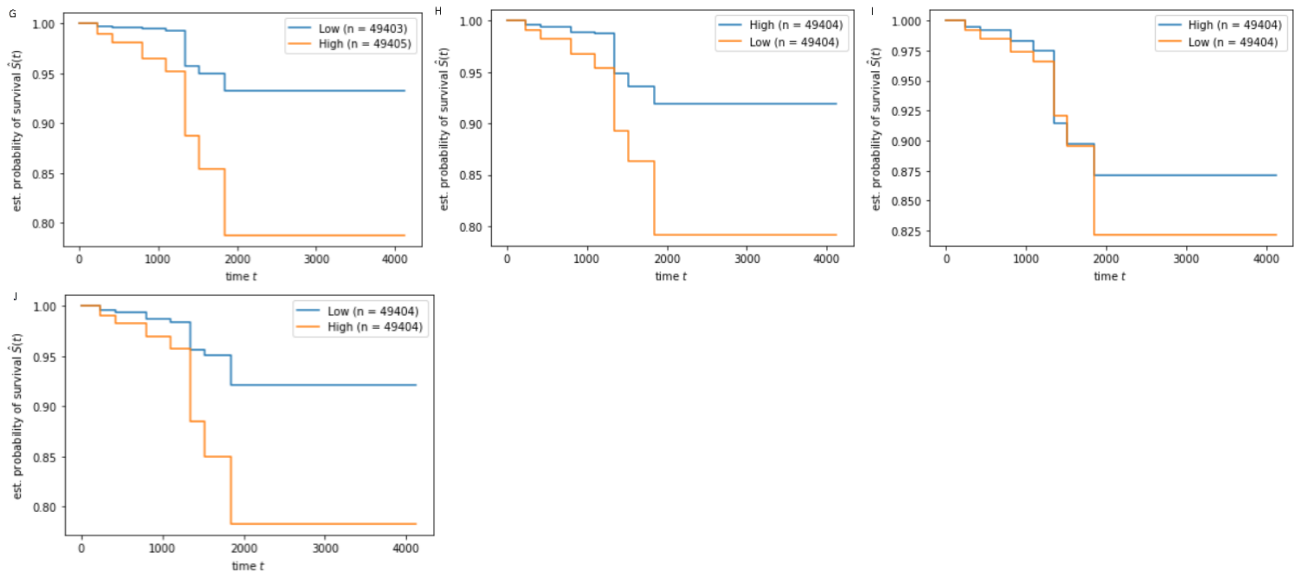

*Figure 5 Kaplan Meier samples stratification into low or high-risk survival group according to the top 10 most significant features*

## 4. Discussion

Colon cancer stages classification and prognostic provide the information about the tumor size, extent of spread of the disease to the nearest lymph nodes and metastasis to a distance site which are necessary for clinical treatment and survival. Our fused features from histopathological whole slide images and functional genomics provide an enhanced complementary model for effective stages prediction. The resulting fused features' latent space vectors aggregate microenvironment of tumors and molecular signals in both datasets.

Although our proposed ensemble deep neural network model shows an improvement in cancer stages prediction and stratification accuracy, still several constraints are encountered during the study. First, only three genomics datasets are considered, extending the study to include other genomics and transcriptional dataset such as somatic mutation (SM), Copy number variation (CNV), and Reverse Phase Protein Array (RPPA) data expression could result into higher accuracy in prediction and risk stratification. Also, the algorithm for integrating both features might have effect on the performance of the classifier, hence consideration and implementation of other integration algorithms are areas to be considered in future studies. Other notable limitation is the absence of independent datasets that involve both image and genomics data from the same samples, to verify and test our proposed framework. This will also be considered in future study. Stratification of samples into low and high-risk survival was based on the median values of each extracted features which can be considered a deterministic approach, this could be compare with a stochastic approach of stratification in a future study.

Other cancer research areas that might benefit from our proposed framework include the classification of cancer of unknown, discernment of tumor microenvironment and microsatellite instability.

In conclusion, our proposed model based on fusion of features extracted from histopathological images and genomics (mRNA, miRNA, DNA Methylation) for colon cancer stages classification and survival risk stratification result in improvement in classification accuracy and viable survival stratification.

This research work was able to confirmed the use of genomics sequencing datasets as an established gold standard for cancer staging, nevertheless, the strength of our proposed framework is evidence in the clearly and the superiority of the t-SNE visual mapping of the integrated/fused features into the four cancer stages as shown in Figure 2C compare with the individual t-SNE visual mapping of genomics Figure 2A and the visual t-SNE visual mapping of images Figure 2B in which the stages are not clearly delineated. This evidence of integrated features superiority over separate features from genomics and images was confirmed by the improved accuracy result as presented in the results section.